# Classification of Colorectal Cancer Polyps via Transfer Learning and Vision-Based Tactile Sensing

Nethra Venkatayogi[1], Ozdemir Can Kara[2], Jeff Bonyun[2], Naruhiko Ikoma[3] and Farshid Alambeigi[2]
[1]Department of Biomedical Engineering, University of Texas at Austin, Austin, TX, USA
[2]Walker Department of Mechanical Engineering, University of Texas at Austin, Austin, TX, USA
[3]Department of Surgical Oncology, MD Anderson Cancer Center, Houston, TX, USA
farshid.alambeigi@austin.utexas.edu

*Abstract*—In this study, to address the current high early-detection miss rate of colorectal cancer (CRC) polyps, we explore the potentials of utilizing transfer learning and machine learning (ML) classifiers to precisely and sensitively classify the type of CRC polyps. Instead of using the common colonoscopic images, we applied three different ML algorithms on the 3D textural image outputs of a unique vision-based surface tactile sensor (VS-TS). To collect realistic textural images of CRC polyps for training the utilized ML classifiers and evaluating their performance, we first designed and additively manufactured 48 types of realistic polyp phantoms with different hardness, type, and textures. Next, the performance of the used three ML algorithms in classifying the type of fabricated polyps was quantitatively evaluated using various statistical metrics.

## I. INTRODUCTION

Colorectal Cancer (CRC) is one of the leading causes of cancer related incidences and mortality worldwide [1]. In 2020, CRC was the third most prevalent cancer type, accounting for 1.9 million new cases and the second leading cause of cancer deaths (935,000 deaths) [1]. Early detection of pre-cancerous lesions (i.e., polyps) via colonoscopy can increase the survival rate of CRC patients to almost 90% [2]. In a colonoscopy screening, morphological characteristics (i.e., shape, size, and texture) of CRC polyps are known to be essential for the classification of different polyp types (Fig. 1) [3], [4]. Nevertheless, due to the high degree of variation in polyp morphology across patients, decision-making based on these characteristics can be very complex and evaluator-dependent [5]. Of note, this has resulted in an early-detection miss rate (EDMR) of approximately 20%-30%, which demands the development of new diagnostic tools to reduce EDMR [6], [7].

Computer aided diagnostics (CAD) using various machine learning (ML) classifiers have shown promising results in improving the detection and characterization of various polyps and particularly reducing the EDMR of CRC polyps [8]–[10]. Specifically, support vector machine (SVM), k-nearest

*Research reported in this publication was supported by the University of Texas at Austin and MD Anderson Cancer Center Pilot Seed Grant.

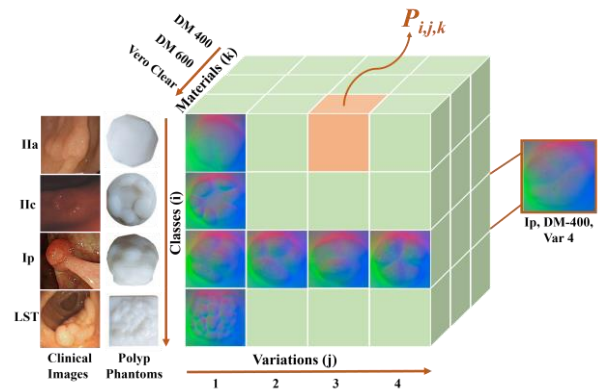

Fig. 1. A conceptual three-dimensional tensor showing the dimensions of the used image dataset collected by the VS-TS on the fabricated polyp phantoms.

neighbors (k-NN), ensemble methods, and convolutional neural networks (CNN) have been at the forefront of endoscopy-based polyp detection and classification algorithms [8]–[10]. More details about ML-based CAD under visual feedback of colonoscopy screening can be found in this paper [8].

Access to a large dataset of images is crucial to developing successful and robust ML classifiers. This requirement is one of the main barriers to performing ML-based CADs, as obtaining medical data and patient records can be very challenging and time-intensive [6]. Further, for the case of CRC polyps, as shown in Fig. 1, a high-degree of variation in CRC polyps' type, shape, texture, and size, which may also vary among cancer patients, would add to the difficulty of preparing a rich and high-quality image dataset [3], [4]. To overcome this barrier, transfer learning has shown great potential in medical CAD and particularly the detection and classification of polyps in endoscopy images [11], [12]. In transfer learning, the learned capabilities for general image recognition subtasks (e.g. pattern identification and feature extraction) can be transferred from a large and exhaustively-trained model and used as the starting point for training a new model on a much smaller dataset [13]. Many neural networks have been trained on large, natural image datasets, such as ImageNet, which features over 1 million images spanning across 1000 categories and has been previously used for texture/pattern identification and classification tasks [14], [15]. Aside from data access concerns, colonoscopic screening is susceptible to issues such as camera occlusions, which can result in overlooked polyps at colon flexures. As a result, in such cases, due to low-quality video footage of the colonoscope, ML-based CAD of CRC polyps may fail and, therefore, cannot be beneficial in reducing the EDMR [6], [7].

To collectively address the above-mentioned challenges in performing an ML-based CAD and reduce the EDMR of CRC polyps, we have developed a vision-based surface tactile sensor (VS-TS) (shown in Fig. 2) that can provide high-resolution 3D textural images of CRC polyps [16], [17]. Fig. 1 illustrates exemplary textural images of this VS-TS applied on various CRC polyp phantoms. As our main contributions in this paper, through the use of 3D textural images of the VS-TS, we explore the potentials of transfer learning to address the medical data access issue and precisely and sensitively classify CRC polyps using three different ML classifiers. To collect realistic textural images of CRC polyps for training the ML models and evaluating their performance, we designed and additively manufactured 48 types of realistic polyp phantoms with different hardness, type, and textures. The performance of the ML models in classifying the type of fabricated polyps, solely based on textural images of the VS-TS, was quantitatively evaluated and compared using various statistical metrics (i.e., accuracy, precision, and sensitivity).

## II. MATERIALS AND METHODS

### A. Vision-based Surface Tactile Sensor (VS-TS)

As shown in Fig. 2, VS-TS consists of (i) a dome-shape deformable silicone membrane (a soft transparent platinum cure two-part silicone, P-565, Silicones Inc.) that directly interacts with the polyp phantoms; (ii) an optics module (Arducam 1/4 inch 5 MP camera) that faces toward the gel layer and captures the deformation of the gel layer; (iii) a transparent acrylic layer that supports the gel layer; (iv) an array of Red, Green and Blue LEDs (WL-SMTD Mono-Color 150141RS63130, 150141GS63130, 150141BS63130) to provide internal illumination; and (v) a rigid frame supports the deformable membrane and Red-Green-Blue LEDs.

### B. Realistic Polyp Phantoms

As conceptually shown in the three-dimensional tensor of Fig. 1, to create our image dataset, we designed and fabricated realistic CRC polyp phantoms by varying three indices (i, j, k) of this tensor. Of note, in every element $P_{i,j,k}$ of this tensor, index $i$ represents the types IIa, IIc, Ip, LST of CRC polyp according to the Paris classification [18] (called T1, T2, T3 and T4, throughout the paper, respectively); index $j$ denotes four different geometrical variations within one type; and index $k$ shows the changes in the stiffness/hardness of the fabricated polyps by choosing three different materials including Agilus DM400 (Shore A 1-2), Agilus DM600 (Shore A 30-40), and Vero Pure White (Shore D 83-86) [19]. Notably, all 4×4×3 = 48 variations of these realistic, high-resolution polyps were printed with the J750 Digital Anatomy Printer (Stratasys, Ltd) and the mentioned materials.

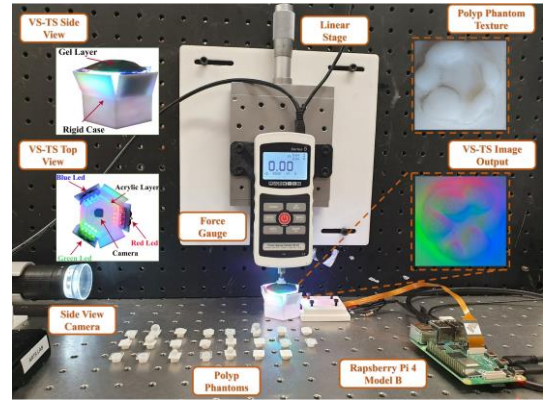

Fig. 2. Experimental setup used for collecting the CRC polyps images.

### C. Experimental Setup and Procedure

Fig. 2 shows the experimental setup utilized to collect textural images using the VS-TS and fabricated polyps. To conduct experiments, each polyp was attached to the force gauge (Mark-10 Series 5, Mark-10 Corporation) through their threaded connections. A linear stage (M-UMR12.40, Newport) was used to push the polyp onto the VS-TS until 8 N of force, in which we could collect high-resolution textural images of tumors using the VS-TS. Of note, these steps were repeated for all 48 unique polyp phantoms. Images from the conducted experiment were then cropped to exhibit only the polyp of interest, and resized to 224×224 pixels. To increase the amount of data, we utilized data augmentation techniques [11], [20] by rotating each of the images in the tensor (i.e., element $P_{i,j,k}$ in Fig. 1) by 90°, 180°, 270°, and then vertically flipping each rotated image and the original image. This yielded 7 additional images for each of our 48 original images, resulting in a total of 384 samples for our classifiers.

### D. ML Classifiers and Classification Procedure

The classifiers used in this experiment were support vector machine (SVM) [21], Residual Neural Network [22] with 18 layers (ResNet-18) trained from random initial weights ($ResNet_1$), and ResNet-18 pre-trained on ImageNet ($ResNet_2$). SVM is a common linear classifier for image classification where hyperplanes are constructed to separate the data with the goal of maximizing the distance between the data of various classes [21]. ResNets are CNNs that are designed to reduce the effects of vanishing gradients in neural networks. This architecture contains residual blocks and shortcut connections, which alleviate the common degradation problem in deep neural networks by facilitating network layers to learn an identity mapping during training [22].

The full dataset was split into training, validation, and testing datasets in a 2:1:1 ratio, creating 12 folds of cross validation. Both ResNet models were trained with a training set of 192 images, validated with a set of 96 images, and tested with a testing set of 96 images. For *SVM*, we trained with a training set of 192 images and tested with a testing set of 96 images using a polynomial kernel function and a grid search over the regularization parameter (C) from a range of 0.01,0.1,1,10,100,1000. A C value of 1.0 was chosen consistently across all folds. For the *ResNet-18* models, a stochastic gradient descent (SGD) optimizer was selected, and early stopping based on validation loss was enabled for all runs to avoid over-fitting. We trained the *ResNet$_1$* model with a range of 49 to 50 epochs for each fold and a learning rate of 0.001. For *ResNet$_2$*, the model was initialized with pre-trained weights and fine-tuned, and the model was trained for a range of 19-20 epochs for each fold. A lower learning rate of 0.0001 was chosen to prevent the loss of the transferred knowledge when fine-tuning [23].

TABLE I
EVALUATION METRICS OF THE UTILIZED CLASSIFIERS

|                 | SVM    | ResNet$_1$ | ResNet$_2$ |
|-----------------|--------|------------|------------|
| Validation Acc. | N/A    | 54.95%     | 92.88%     |
| Test Acc.       | 53.30% | 54.95%     | 91.93%     |
| Sensitivity     | 53.30% | 54.95%     | 91.27%     |
| Precision       | 54.42% | 56.27%     | 92.06%     |

*E. Evaluation Metrics*

To evaluate the performance of the used classifiers, we utilize accuracy (*A*), sensitivity (*S*), and precision (*P*). Additionally, it is clinically important to also consider metrics respective to each class pair and analyze *pair-wise sensitivity* ($C_S^{i,j}$), and *pair-wise precision* ($C_P^{i,j}$), which can be calculated as below:

$$C_S^{i,j} = \frac{|y_i \cap \hat{y}_j|}{|y_i|} \qquad C_P^{i,j} = \frac{|y_i \cap \hat{y}_j|}{|\hat{y}_j|} \qquad (1)$$

where *y* is the set of true labels, $\hat{y}$ is the set of predicted labels, and *i* and *j* refer to the predicted and true class labels, respectively. $C_S^{i,j}$ measures the model's ability to correctly label polyps and $C_P^{i,j}$ measures the probability of how correct that prediction is.

### III. RESULTS AND DISCUSSION

Validation and test accuracy were recorded and averaged across folds for all models. Sensitivity, precision, and pair-wise sensitivity and precision normalized confusion matrices were calculated through the predictions from the test set (Fig. 3). Table I reports the average validation and testing metrics. As seen, *ResNet$_2$* achieves validation and test accuracy of around 92%, obtaining better metrics than both *ResNet$_1$* and SVM. Compared with existing classifiers utilizing conventional colonoscopy images, *ResNet$_2$*'s metrics were lower than Wang et al.'s classifier [24] that utilized a ResNet-50 model and achieved an accuracy of around 95%. However, it should be noted that their dataset was considerably larger, and they did not use a tactile sensing device. Our *ResNet$_2$* model achieved better metrics than Ribeiro et al.'s MatConvNet implementation of the Slow CNN model [11], which achieved an accuracy of 91.42% as well as Patel et al.'s study [6] that experimented with ResNet-50 and other pre-trained architectures, which achieved accuracies around 80%. Of note, these studies also utilized colonoscopic visual images, not a VS-TS textural output.

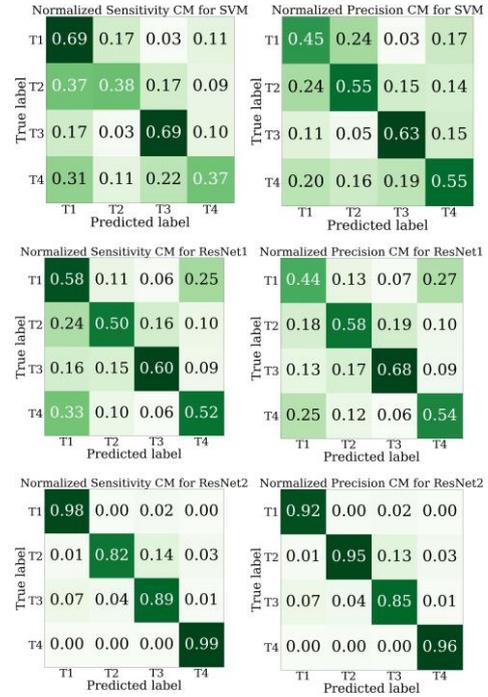

Fig. 3. Pair-wise Sensitivity (first column) and Precision (second column) confusion matrices of the utilized ML classifiers.

As shown in Fig. 3, the *Resnet$_2$* model also achieved consistent strong performance in the classification of each of the four classes (with sensitivity $\geq$ 0.82, precision $\geq$ 0.85, in every case). The *Resnet$_2$* model had high values for both T1 and T4 polyps $C_S^{T1,T1} = .98$ and $C_S^{T4,T4} = 0.92$ as well as their precision counterparts with $C_P^{T1,T1} = 0.92$ and $C_P^{T4,T4} = 0.96$. Since T1 and T4 polyps are generally known as flat polyps, which are the most difficult to diagnose during screening, this exhibits the model's capability in distinguishing challenging cases [25]. Neither the SVM nor the random-initialization *ResNet$_1$* models performed well on any of our metrics. The success of the pretrained *ResNet$_2$* indicates that the high dimensionality of the data (224 ×224 images × 3 RGB channels = 150,528 values in each sample) combined with the small sample size (training sets of 192 samples) exceeded the ability of a non-pretrained network to extract the signal from the noise. Notably, this emphasizes the value of transfer learning in small medical datasets [11].

To conclude, the obtained results in this study proved, (i) better classifiers can be achieved by transfer learning from pre-trained neural network architectures trained on general, and non-medical images; and (ii) the textural information obtained from the VS-TS sensor can be used by such classifiers to deliver high accuracy in polyp classification, competitive with or better than 2D image classifiers. Future work will primarily focus on increasing the size of our dataset, simulating more variations in realistic polyps and then transition to the analysis of the efficiency of our model when trained to recognize common morphological features in video frames of real-life samples and focus on testing of our system on clinical data.